\documentclass{article}
\usepackage{spconf,amsmath,amssymb,graphicx,booktabs,tikz,microtype,xurl}
\usepackage{newtxtext,newtxmath}
\usepackage{cleveref}
\usepackage{pgfplots} \pgfplotsset{compat=1.18}
\usepackage[T1]{fontenc}
\usepackage[table]{xcolor}
\usetikzlibrary{arrows.meta,calc}
\usepackage{anyfontsize,balance,etoolbox,needspace}
\apptocmd{\thebibliography}{\tolerance=5000\emergencystretch=3em}{}{}

\definecolor{resultshade}{HTML}{EDF5F5}

\newcommand{\best}[1]{\textbf{#1}}
\newcommand{\up}{\ensuremath{\uparrow}}
\newcommand{\dn}{\ensuremath{\downarrow}}

\title{Adapting Personalized Speech Enhancement for\\Low-Latency Audio-Visual Target-Speaker Extraction}
\name{Rayhan Rashed$^{\star\clubsuit}$ \qquad Senja Filipi$^{\star}$ \qquad Ross Cutler$^{\star}$}
\address{$^{\star}$Microsoft \qquad $^{\clubsuit}$University of Michigan}
\begin{document}
\ninept
\raggedbottom
\emergencystretch=.75em
\maketitle

\begin{abstract}
Online audio-visual target-speaker extraction aims to remove competing voices while preserving speech quality and bounding lookahead. Existing extractors are built and evaluated for separation on synthetic mixtures, leaving listening quality and meeting behavior largely untested. We introduce Audio-Visual Personalized Voice Quality Enhancement (AV-PVQE), which approaches these requirements from the other direction. We start from a personalized speech enhancement model that reconstructs a requested voice at high quality but confuses the target in 46\% of two-speaker mixtures despite clean enrollment. Adding mouth features at its speaker-conditioning input and jointly fine-tuning the visual and reconstruction networks reduces this rate to 1.6\%, with no future frames and 20 ms of algorithmic delay. Compared with an online autoregressive audio-visual extractor, AV-PVQE yields separation gains on two synthetic benchmarks and larger gains on recorded meetings, and keeps its advantage on excerpts with more speakers than the fine-tuning mixtures. In personalized P.835 listening tests on two meeting corpora, it improves overall quality over this extractor by 0.57 and 0.63 MOS, with similar mean rating relative to the starting model. Preservation and rejection tests show that it keeps the target intact when no competing voice is present and suppresses competing speech when the target is absent.
\end{abstract}

\vspace{-4pt} 
\section{Introduction}
\vspace{-4pt}
Target-speaker extraction (TSE) recovers a requested speaker from audio containing competing voices. In a video call, this can keep nearby conversations out of the participant's audio. An online extractor works as the recording arrives, with little added delay, while the participant talks over others, talks alone, and falls silent as others keep talking. A useful extractor keeps the participant's voice intact in the first two cases and removes the remaining voices in the third.
 
Extraction needs an auxiliary cue that tells the network which speaker to keep. The cue is commonly a pre-recorded sample of the target's voice \cite{wang2019voicefilter,eskimez2022personalized} or a recording of the target's face, which the network matches to the synchronized speech \cite{ephrat2018looking,pan2022usev}; some models use both \cite{wu2024unified}. Pan et al.\ \cite{pan2025avase} apply the visual cue online, in a model referred to here as AVASE. It runs frame by frame, and a lightweight acoustic encoder feeds its own previously extracted frames back as a second cue. It is trained from scratch on simulated two-speaker mixtures and evaluated on mixtures of the same kind, with separation, speech-quality, and intelligibility scores. 

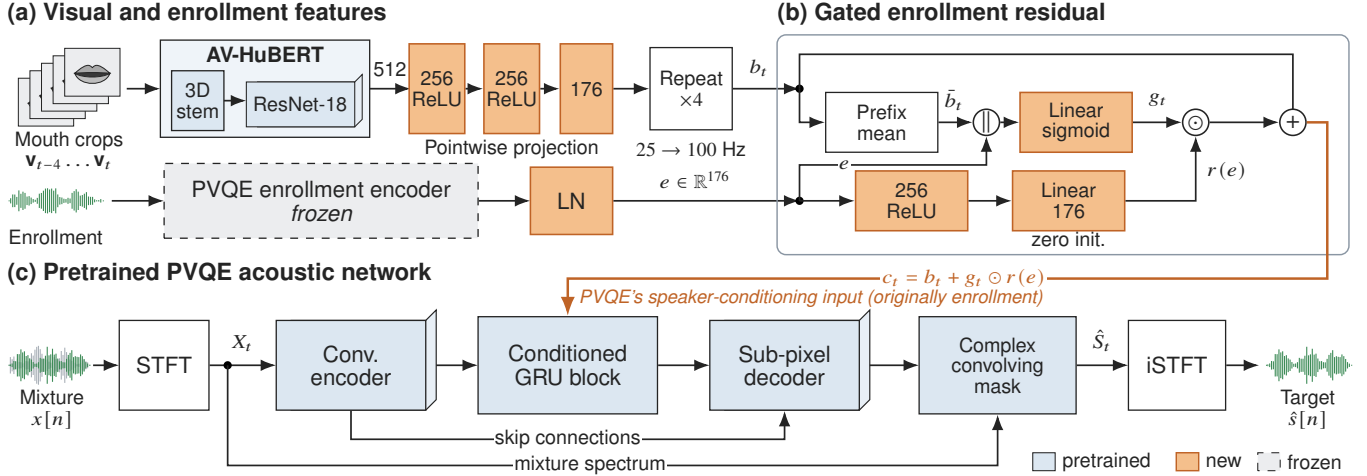
\begin{figure*}[t]
\centering
\resizebox{\textwidth}{!}{%
\begingroup%
\definecolor{avink}{HTML}{242424}%
\definecolor{avpre}{HTML}{DCE7F1}%
\definecolor{avnew}{HTML}{F4BF8C}%
\definecolor{avnewd}{HTML}{BF6326}%
\definecolor{avgray}{HTML}{ECEDEF}%
\definecolor{avline}{HTML}{8A96A3}%
\definecolor{avtgt}{HTML}{3D8B57}%
\definecolor{avint}{HTML}{A9B0B8}%
\begin{tikzpicture}[
  x=1mm,y=1mm,>=Latex,
  font=\sffamily\fontsize{8.5}{9.5}\selectfont,
  text=avink,inner sep=0pt,
  flow/.style={-Latex,draw=avink,line width=.65pt},
  wire/.style={draw=avink,line width=.65pt},
  module/.style={draw=avink,line width=.5pt,align=center,inner sep=1pt},
  pre/.style={module,fill=avpre},
  new/.style={module,fill=avnew,draw=avnewd,line width=.6pt},
  opbox/.style={module,fill=white},
  small/.style={font=\sffamily\fontsize{7.5}{8.2}\selectfont},
  note/.style={font=\sffamily\itshape\fontsize{7}{7.8}\selectfont},
  panel/.style={anchor=west,font=\sffamily\bfseries\fontsize{9}{10}\selectfont},
  op/.style={circle,draw=avink,fill=white,line width=.6pt,
    minimum size=3.5mm,font=\fontsize{9}{9}\selectfont}
]
\path[use as bounding box] (0,-7) rectangle (180,57);

\newcommand{\avwave}[7]{%
  \pgfmathsetseed{#6}%
  \foreach \i in {0,...,46}{%
    \pgfmathsetmacro{\tt}{\i/46}%
    \pgfmathsetmacro{\ee}{(0.16+0.84*abs(sin(540*\tt+#7))^1.5)*sin(180*\tt)^0.35}%
    \pgfmathsetmacro{\aa}{#4*\ee*(0.35+0.65*rnd)}%
    \draw[#5,line width=.42pt] ({#1+\tt*#3},{#2-\aa}) -- ({#1+\tt*#3},{#2+\aa});}}
\newcommand{\avmouth}[1]{%
  \filldraw[fill=avgray,draw=avink,line width=.35pt] (0,0) rectangle (7.4,6.4);
  \fill[black!78] (1.35,3.05) .. controls (2.4,3.2) and (5.0,3.2) .. (6.05,3.05)
    .. controls (5.0,{3.05-#1}) and (2.4,{3.05-#1}) .. cycle;
  \filldraw[fill=black!34,draw=avink,line width=.3pt] (1.35,3.05)
    .. controls (2.1,3.75) and (2.85,4.25) .. (3.3,4.1) .. controls (3.5,4.0) .. (3.7,3.82)
    .. controls (3.9,4.0) .. (4.1,4.1) .. controls (4.55,4.25) and (5.3,3.75) .. (6.05,3.05)
    .. controls (5.0,3.2) and (2.4,3.2) .. cycle;
  \filldraw[fill=black!26,draw=avink,line width=.3pt] (1.35,3.05)
    .. controls (2.4,{3.05-#1}) and (5.0,{3.05-#1}) .. (6.05,3.05)
    .. controls (5.4,{1.55-0.45*#1}) and (2.0,{1.55-0.45*#1}) .. cycle;}

\node[panel] at (0,55) {(a) Visual and enrollment features};
\node[panel] at (103,55) {(b) Gated enrollment residual};
\node[panel] at (0,20.2) {(c) Pretrained PVQE acoustic network};

\def\Yv{45}
\foreach \i/\o in {0/0.25,1/0.55,2/0.80,3/0.40,4/0.65}{
  \begin{scope}[shift={(1.5+1.55*\i,40+1.0*\i)}]\avmouth{\o}\end{scope}}
\node[small,align=center] at (8.3,36.7) {Mouth crops\\$\textbf{v}_{t-4}\ldots\textbf{v}_t$};
\draw[flow] (16,\Yv) -- (20.5,\Yv);
\draw[fill=avpre!45,draw=avink,line width=.5pt] (20.5,38.5) rectangle (48.5,51.5);
\node[font=\sffamily\bfseries\fontsize{8}{9}\selectfont] at (34.5,49.6) {AV-HuBERT};
\node[pre,minimum width=7mm,minimum height=8mm,small] (stem) at (25.5,43.4) {3D\\stem};
\draw[flow] (29,43.4) -- (32,43.4);
\draw[fill=avpre!70,draw=avink,line width=.4pt] (33.5,39.9) -- (46.5,39.9) -- (47.5,40.9) -- (47.5,46.9) -- (46.5,45.9) -- (33.5,45.9) -- cycle;
\draw[fill=avpre,draw=avink,line width=.4pt] (32,39.9) rectangle (46.5,45.9);
\node[small] at (39.2,42.9) {ResNet-18};
\draw[flow] (48.5,\Yv) -- (54,\Yv);
\node[small] at (51.2,\Yv+2.6) {512};
\node[new,minimum width=7mm,minimum height=12mm,small] (p1) at (57.5,\Yv) {256\\ReLU};
\node[new,minimum width=7mm,minimum height=12mm,small] (p2) at (67.5,\Yv) {256\\ReLU};
\node[new,minimum width=7mm,minimum height=12mm,small] (p3) at (77.5,\Yv) {176};
\draw[flow] (p1.east) -- (p2.west);
\draw[flow] (p2.east) -- (p3.west);
\node[small] at (67.5,36.9) {Pointwise projection};
\node[opbox,minimum width=11mm,minimum height=11mm,small] (hold) at (91.5,\Yv) {Repeat\\$\times4$};
\draw[flow] (p3.east) -- (hold.west);
\node[small] at (91.5,36.9) {$25\rightarrow100$ Hz};
\draw[flow] (hold.east) -- (106,\Yv);
\node[small,fill=white,inner sep=.3pt] at (100.6,\Yv+2.8) {$b_t$};

\def\Ye{30}
\avwave{0.2}{\Ye}{12.6}{2.15}{avtgt}{23}{35}
\node[small] at (6.5,25.1) {Enrollment};
\draw[flow] (14,\Ye) -- (21,\Ye);
\node[module,dashed,fill=avgray,minimum width=42mm,minimum height=10mm] (enroll) at (42,\Ye) {PVQE enrollment encoder\\\textit{frozen}};
\draw[flow] (enroll.east) -- (70,\Ye);
\node[new,minimum width=11mm,minimum height=9mm] (ln) at (75.5,\Ye) {LN};
\draw[flow] (ln.east) -- (106,\Ye);
\node[small,fill=white,inner sep=.4pt] at (92,\Ye+2.8) {$e\in\mathbb{R}^{176}$};

\draw[draw=avline,line width=.55pt,rounded corners=1.2mm] (103,22.8) rectangle (179.5,52.2);
\def\Ym{40.5}
\fill (106,\Yv) circle (.55mm);
\draw[wire] (106,\Yv) -- (106,49.6) -- (172,49.6) -- (172,{\Ym+1.75});
\node[opbox,minimum width=15mm,minimum height=8mm,small] (mean) at (117,\Ym) {Prefix\\mean};
\draw[flow] (106,\Yv) -- (106,\Ym) -- (mean.west);
\node[op] (cat) at (131,\Ym) {$\Vert$};
\draw[flow] (mean.east) -- (cat.west);
\node[small] at (126.7,\Ym+3) {$\bar b_t$};
\node[new,minimum width=15mm,minimum height=8mm,small] (gate) at (143,\Ym) {Linear\\sigmoid};
\draw[flow] (cat.east) -- (gate.west);
\node[op] (mul) at (159,\Ym) {$\odot$};
\draw[flow] (gate.east) -- (mul.west);
\node[small] at (154,\Ym+3) {$g_t$};
\node[op] (add) at (172,\Ym) {$+$};
\draw[flow] (mul.east) -- (add.west);
\fill (106,\Ye) circle (.55mm);
\draw[flow] (106,\Ye) -- (106,34.8) -- (131,34.8) -- (cat.south);
\node[small,fill=white,inner sep=.4pt] at (112,34.8) {$e$};
\node[new,minimum width=15mm,minimum height=8mm,small] (r1) at (121,\Ye) {256\\ReLU};
\node[new,minimum width=15mm,minimum height=8mm,small] (r2) at (142,\Ye) {Linear\\176};
\draw[flow] (106,\Ye) -- (r1.west);
\draw[flow] (r1.east) -- (r2.west);
\draw[flow] (r2.east) -- (159,\Ye) -- (mul.south);
\node[small,anchor=west] at (160.5,34) {$r(e)$};
\node[small] at (142,24.9) {zero init.};

\draw[-Latex,draw=avnewd,line width=.95pt] (add.east) -- (176.6,\Ym) -- (176.6,19.6) -- (75,19.6) -- (75,14.6);
\node[small,fill=white,inner sep=.5pt,text=avnewd] at (128,19.6) {$c_t=b_t+g_t\odot r(e)$};
\node[note,anchor=west,text=avnewd] at (76.4,16.95) {PVQE's speaker-conditioning input (originally enrollment)};

\def\Yc{8}
\avwave{0.2}{\Yc}{10.8}{2.9}{avint}{11}{80}
\avwave{0.2}{\Yc}{10.8}{2.9}{avtgt}{7}{0}
\node[small,align=center] at (6,2.1) {Mixture\\$x[n]$};
\draw[flow] (11.5,\Yc) -- (15,\Yc);
\node[opbox,minimum width=12mm,minimum height=12mm] (stft) at (21,\Yc) {STFT};
\draw[flow] (stft.east) -- (36,\Yc);
\fill (29.5,\Yc) circle (.55mm);
\node[small] at (31.5,\Yc+3) {$X_t$};
\draw[fill=avpre!65,draw=avink,line width=.45pt] (38,2) -- (56,2) -- (57.4,3.4) -- (57.4,15.4) -- (56,14) -- (38,14) -- cycle;
\node[pre,minimum width=20mm,minimum height=12mm] (enc) at (46,\Yc) {Conv.\\encoder};
\draw[flow] (57.4,\Yc) -- (63,\Yc);
\node[pre,minimum width=24mm,minimum height=13mm] (rnn) at (75,\Yc) {Conditioned\\GRU block};
\draw[flow] (rnn.east) -- (94,\Yc);
\draw[fill=avpre!65,draw=avink,line width=.45pt] (96,2) -- (114,2) -- (115.4,3.4) -- (115.4,15.4) -- (114,14) -- (96,14) -- cycle;
\node[pre,minimum width=20mm,minimum height=12mm] (dec) at (104,\Yc) {Sub-pixel\\decoder};
\draw[flow] (115.4,\Yc) -- (122,\Yc);
\node[pre,minimum width=21mm,minimum height=13mm,small] (ccm) at (132.5,\Yc) {Complex\\convolving\\mask};
\draw[flow] (ccm.east) -- (150,\Yc);
\node[small] at (146.5,\Yc+3.3) {$\hat S_t$};
\node[opbox,minimum width=13mm,minimum height=12mm] (istft) at (156.5,\Yc) {iSTFT};
\draw[flow] (istft.east) -- (167.5,\Yc);
\avwave{168.4}{\Yc}{11.4}{2.9}{avtgt}{7}{0}
\node[small,align=center] at (174,2.1) {Target\\$\hat s[n]$};
\draw[flow] (enc.south) -- (46,-1.9) -- (104,-1.9) -- (dec.south);
\node[small,fill=white,inner sep=.4pt] at (75,-1.9) {skip connections};
\draw[flow] (29.5,\Yc) -- (29.5,-5.4) -- (132.5,-5.4) -- (ccm.south);
\node[small,fill=white,inner sep=.4pt] at (78,-5.4) {mixture spectrum};

\begin{scope}[shift={(137.2,-4.9)}]
  \filldraw[fill=avpre,draw=avink,line width=.4pt] (0,-1.2) rectangle (3,1.2);
  \node[small,anchor=west] at (3.9,0) {pretrained};
  \filldraw[fill=avnew,draw=avnewd,line width=.55pt] (19.2,-1.2) rectangle (22.2,1.2);
  \node[small,anchor=west] at (23.1,0) {new};
  \filldraw[fill=avgray,draw=avink,dashed,line width=.4pt] (30.1,-1.2) rectangle (33.1,1.2);
  \node[small,anchor=west] at (34,0) {frozen};
\end{scope}
\end{tikzpicture}%
\endgroup
}
\vspace{-8pt}
\caption{AV-PVQE: new layers (orange) connect two pretrained networks (blue), and $c_t$ enters PVQE's speaker-conditioning input. All trainable blocks except the enrollment encoder are fine-tuned.}
\label{fig:architecture}
\vspace{-5pt}
\end{figure*}
 
Speech enhancement pursues a different objective, improving listening quality. DeepVQE \cite{ristea2023deepvqe}, for example, handles noise suppression, acoustic echo cancellation, and dereverberation, usually separate tasks, in one common network. It outperformed the Deep Noise Suppression Challenge \cite{dubey2023dns} winners in subjective ratings, and is used in Microsoft Teams. Personalized DeepVQE (PVQE) \cite{parnamaa2024personalized} extends DeepVQE with an enrollment recording that guides the model to retain the enrolled speaker and suppress the rest. Conceptually this is close to the TSE task, since both reconstruct one requested voice and remove the rest. In PVQE, this request enters through a speaker-conditioning input that carries a single vector specifying the voice to retain. We route the visual cue through this same input. In this paper, we ask whether an online audio-visual TSE model built  this way from PVQE selects the target as reliably as a model trained from scratch for the task, while keeping PVQE's listening quality.
 
Applying PVQE directly to two-speaker mixtures exposes a selection failure. On synthetic mixtures from LRS3, a corpus of TED-talk videos \cite{afouras2018lrs3}, its output is closer to the competing speaker than to the target in 46\% of cases despite clean enrollment. Extracting the wrong speaker is known as target confusion \cite{zhao2022confusion}. Yet DNSMOS, a neural predictor of speech, background, and overall quality ratings \cite{reddy2022dnsmosp835}, scores PVQE's outputs above AVASE's (\Cref{tab:extraction}). It receives only the output audio, with no enrollment, so its scores cannot show which speaker was recovered.
 
We introduce Audio-Visual Personalized Voice Quality Enhancement (AV-PVQE), which adds mouth features at PVQE's speaker-conditioning input and combines them with enrollment. We jointly fine-tune the visual conditioning and PVQE's reconstruction network on two-speaker mixtures, using only current and earlier video frames. Two earlier systems are closest to this design. AV-E3Net takes the architecture of E3Net, a personalized speech enhancement model, and trains it with mouth features in place of enrollment \cite{zhu2023ave3net}; AV-PVQE keeps both cues, as other extractors do \cite{wu2024unified}, and starts from PVQE's trained weights. Plug-and-Steer keeps a pretrained separator frozen and steers the target to one of its output channels \cite{kwak2026plugsteer}. PVQE produces a single output, so when it keeps the wrong voice there is no other channel to select; fine-tuning instead changes which voice the network reconstructs.
 
We compare AV-PVQE with AVASE and PVQE on synthetic two-speaker mixtures from LRS3 and VoxCeleb2, a corpus of interview videos \cite{chung2018vox2}, and, without further fine-tuning, on three meeting corpora: AMI \cite{carletta2005ami} (summed headset channels), MCoRec \cite{nguyen2025mcorec,mcorec2026data} (in-person conversations at a central microphone), and MTM (Teams meetings with two to four active speakers). We measure target recovery against source references, and we test whether AV-PVQE preserves the target when it speaks alone and suppresses competing speech when it is silent, following preservation and rejection tests from personalized and audio-visual extraction \cite{parnamaa2024personalized,eskimez2022personalized,pan2022continuity,pan2022usev}. Listeners rate quality in a personalized adaptation of the ITU-T P.835 standard \cite{naderi2021p835,parnamaa2024personalized,dubey2023dns}, hearing the target's voice before each excerpt. \textbf{In summary}, our contributions are:
\begin{enumerate}
\item We propose AV-PVQE, an online audio-visual TSE model built by adding mouth features to the speaker-conditioning input of a pretrained personalized voice quality enhancement model, PVQE and fine-tuning its reconstruction network, with {\textit{zero}} future audio or video frames required.
\item We show that AV-PVQE recovers the target far more reliably than PVQE with similar mean ratings in perceptual quality, and that it exceeds AVASE in target recovery and listener ratings, with larger separation gains on recorded meetings.
\end{enumerate}
\vspace{-6pt} 

\section{AV-PVQE model}
\vspace{-3pt} 
\noindent\textbf{Adapting the enhancement network:}
Given an audio mixture $x[n]=s[n]+u[n]$, target video, and an enrollment recording, we estimate the target speech $s[n]$ while removing competing speech and other interference $u[n]$. We initialize the acoustic network from PVQE \cite{parnamaa2024personalized}. Its short-time Fourier transform (STFT), encoder, recurrent layers, and decoder reconstruct the waveform (Fig.~\ref{fig:architecture}). Its frozen enrollment encoder maps a separate target recording to a layer-normalized vector $e\in\mathbb{R}^{176}$, held fixed within each excerpt; PVQE's recurrent block receives $e$ as its speaker-conditioning input.
 
We add mouth features at the speaker-conditioning input using the convolutional encoder of AV-HuBERT \cite{shi2022avhubert}, and project its 512-dimensional features through two 256-dimensional hidden layers with rectified linear units (ReLU) to 176 dimensions. Following AVASE \cite{pan2025avase}, we repeat the visual features to match the audio feature rate. We denote the aligned cue by $b_t$. The visual cue changes with mouth movement, whereas enrollment provides a fixed voice reference. We combine them through a gated enrollment residual:
\begin{equation}
\begin{aligned}
g_t&=\sigma\!\left(W_g[\bar b_t;e]+a_g\right),\\
r(e)&=W_2\operatorname{ReLU}(W_1e+a_1)+a_2,\\
c_t&=b_t+g_t\odot r(e),
\end{aligned}
\label{eq:fusion}
\vspace{-6pt}
\end{equation}
where $\bar b_t=(t+1)^{-1}\sum_{j=0}^{t}b_j$ averages observed features from the excerpt's start, $\sigma$ is the sigmoid, $\odot$ denotes elementwise multiplication, and $r$ has a 256-dimensional hidden layer. We initialize $W_2$ and $a_2$ to zero, so that adding enrollment initially gives $c_t=b_t$, and learn the residual while fine-tuning the visual and reconstruction networks.

\vspace{4pt} 
\noindent\textbf{Fine-tuning for online processing: }
We first train the visual conditioning, then add the enrollment residual and continue training on both corpora. For online operation, we replace whole-excerpt cue aggregation with the running mean in (\ref{eq:fusion}) and remap temporal decoder filters to current and past inputs, initializing newly introduced past taps to zero. We carry audio state across chunks and restrict visual encoding to the current mouth frame and its four predecessors. A final fine-tuning stage updates the visual path, fusion layers, and acoustic network together.
 
Of AV-PVQE's 12.79 million parameters, 0.40 million are new; the rest come from the pretrained AV-HuBERT encoder and PVQE. We process 16 kHz audio with a 20 ms STFT window and a 10 ms hop, and $88\!\times\!88$ grayscale mouth crops at 25 fps. Following DeepVQE \cite{ristea2023deepvqe}, 10 ms of overlap-add delay plus a 10 ms input frame gives 20 ms of algorithmic and buffering delay, meeting the ICASSP 2023 DNS latency limit \cite{dns2023rules}. AV-PVQE uses no future audio or video, unlike a recent real-time audio-visual enhancement model that buffers two future video frames \cite{ma2025raven}. We verified causal streaming operation: changing future video frames left earlier visual outputs bit-exact, and chunked and full-waveform outputs agreed to above 128 dB SI-SNR.
\vspace{-2pt} 
\begin{table*}[t]
\caption{AV-PVQE gives the best target recovery on every corpus. Higher is better. \textbf{Bold} marks the best target-recovery value per corpus; DNSMOS columns are not ranked because the scorer receives no enrollment and cannot identify the recovered speaker (Sec.~\ref{sec:synthetic}).}
\label{tab:extraction}
\centering
\fontsize{9}{10.2}\selectfont
\setlength{\tabcolsep}{3.2pt}
\renewcommand{\arraystretch}{1.08}
\begin{tabular*}{\textwidth}{@{\extracolsep{\fill}}llrrrrrrrr@{}}
\toprule
 & & \multicolumn{4}{c}{Target recovery} & \multicolumn{4}{c}{Predicted quality (DNSMOS)}\\
\cmidrule(lr){3-6}\cmidrule(l){7-10}
Dataset & Model & \shortstack{SI-SNRi\\(dB)\up} & \shortstack{Picked\\(\%)\up} & \shortstack{WAcc\\(\%)\up} & Spk-sim\up & SIG\up & BAK\up & OVRL\up & P.808\up\\
\midrule
LRS3 & AVASE & 8.72 & 96.83 & 88.89 & .890 & 3.508 & 2.803 & 2.564 & 2.915\\
 & PVQE & $-2.31$ & 54.17 & 28.57 & .727 & 3.641 & 3.354 & 2.916 & 3.434\\
 & \best{AV-PVQE (ours)} & \best{9.03} & \best{98.38} & \best{90.91} & \best{.913} & 3.563 & 2.903 & 2.648 & 3.106\\
\cmidrule{2-10}
VoxCeleb2 & AVASE & 3.81 & 85.52 & 75.00 & .843 & 3.683 & 1.887 & 2.146 & 2.807\\
 & PVQE & $-2.97$ & 54.12 & 28.57 & .725 & 3.617 & 3.046 & 2.754 & 3.501\\
 & \best{AV-PVQE (ours)} & \best{5.65} & \best{92.57} & \best{82.35} & \best{.876} & 3.764 & 2.400 & 2.500 & 3.089\\
\cmidrule{2-10}
AMI & AVASE & 2.56 & 71.96 & 50.14 & .864 & 2.593 & 3.879 & 2.195 & 2.858\\
 & PVQE & .40 & 57.01 & 18.51 & .851 & 3.226 & 4.311 & 2.893 & 3.451\\
 & \best{AV-PVQE (ours)} & \best{6.99} & \best{83.18} & \best{68.27} & \best{.885} & 2.947 & 3.967 & 2.481 & 3.065\\
\cmidrule{2-10}
MTM & AVASE & 3.06 & 76.99 & 49.56 & .867 & 2.781 & 3.571 & 2.352 & 3.034\\
 & PVQE & $-.08$ & 51.99 & 8.64 & .854 & 3.587 & 3.939 & 3.125 & 3.402\\
 & \best{AV-PVQE (ours)} & \best{9.64} & \best{90.34} & \best{67.65} & \best{.920} & 3.272 & 3.674 & 2.744 & 3.187\\
\bottomrule
\end{tabular*}
\end{table*}

\section{Experimental setup}
\vspace{-3pt}  \noindent\textbf{Training and baselines:}
We generate two-speaker mixtures (2-spk-mix) from LRS3 and VoxCeleb2 with AVASE's recipe sizes of 40,000 training pairs and 5,000 fixed validation pairs \cite{pan2025avase}, resampling training pairs and temporal excerpts each epoch. The two speakers in a mixture come from the same corpus. Following \cite{pan2025avase}, we draw the target-to-interferer ratio within $-10$ to $10$~dB, and report results for quieter, equal (within $\pm2$~dB), and louder targets. We extract each speaker in turn, using that speaker's video and a separate enrollment utterance. Speakers appearing in the test set for both LRS3 and VoxCeleb2 datasets are unseen in train and validation set, and each test set contains 3,000 fixed pairs\cite{pan2025avase}.

We minimize negative SI-SNRi with Adam \cite{kingma2015adam} at an initial
learning rate of $10^{-3}$. The final fine-tuning stage runs
for 30 epochs with batch size 16. We select the checkpoint
with the highest SI-SNRi on quieter-target validation
mixtures, which occurs at epoch 29.
 
PVQE, the starting point, shares AV-PVQE's reconstruction network and enrollment but receives no video, so it shows what video and fine-tuning add. AVASE represents extraction trained from scratch; we run its released autoregressive checkpoint, which receives video and derives its acoustic cue from its own previous output, with no enrollment. All three systems process the same locally generated test mixtures, so absolute scores are not comparable with AVASE's published results.
 
\begin{table}[t]
\vspace{-8pt}
\caption{AV-PVQE keeps its lead over AVASE on MTM excerpts with more speakers than the two-speaker fine-tuning mixtures. Groups count participants who speak anywhere in an excerpt.}
\label{tab:speakers}
\centering
\fontsize{9}{10.2}\selectfont
\setlength{\tabcolsep}{2.3pt}
\renewcommand{\arraystretch}{1.08}
\begin{tabular}{@{}lrrrr@{}}
\toprule
 & \multicolumn{2}{c}{SI-SNRi (dB)\up} & \multicolumn{2}{c}{Picked (\%)\up}\\
\cmidrule(lr){2-3}\cmidrule(l){4-5}
Mixture group & AVASE & AV-PVQE & AVASE & AV-PVQE\\
\midrule
2-spk-mix & 4.27 & \best{12.16} & 84.10 & \best{93.33}\\
3-spk-mix & 1.92 & \best{7.13} & 72.57 & \best{91.15}\\
4-spk-mix & 0.65 & \best{4.91} & 56.82 & \best{75.00}\\
\bottomrule
\end{tabular}
\vspace{-6pt}
\end{table}

\vspace{6pt}  \noindent\textbf{Meeting evaluation:}
We test six-second excerpts from three meeting corpora without further fine-tuning. In the AMI Meeting Corpus \cite{carletta2005ami}, we sum participant headset channels, so the target's headset is a component of the mixture and serves as its reference, although it also carries crosstalk from nearby speakers. Enrollment uses a separate interval of the target's headset. MTM contains four-participant Teams/Zoom meetings with separate participant channels and their exact sum. We evaluate its target-present excerpts with ten-second target enrollment and group them by the number of participants who speak anywhere in the excerpt.
 
MCoRec contains concurrent in-person conversations \cite{nguyen2025mcorec,mcorec2026data} recorded at a central microphone, which we use as input. The target's worn microphone records the same speech through a different acoustic path, not the target's contribution to the central recording, so it cannot serve as a separation reference. We evaluate MCoRec using word accuracy, speaker similarity, and listening tests.
 
\begin{table*}[t]
\caption{AV-PVQE nearly eliminates PVQE's prolonged target suppression and rejects competing speech far more strongly than AVASE (LRS3). Long TSOS counts events over 1 s; clip-mean attenuation should be near zero for target-only audio and large for competing-only audio. \textbf{Bold} marks the best value per column.}
\label{tab:controls}
\centering
\fontsize{9}{10.2}\selectfont
\renewcommand{\arraystretch}{1.08}
\setlength{\tabcolsep}{5pt}
\begin{tabular*}{\textwidth}{@{\extracolsep{\fill}}lrrrrr@{}}
\toprule
 & \multicolumn{3}{c}{Target-only audio} & \multicolumn{2}{c}{Competing-only audio\up}\\
\cmidrule(lr){2-4}\cmidrule(l){5-6}
Model & Long TSOS\dn & Within $\pm3$ dB (\%)\up & Attenuation (dB) & Static video (dB) & Shifted video (dB)\\
\midrule
AVASE & \best{5} & \best{98.50} & .810 & 2.25 & 8.62\\
PVQE & 142 & 97.42 & .839 & 2.46 & ---\\
AV-PVQE (ours) & 7 & 96.93 & \best{.665} & \best{32.83} & \best{18.28}\\
\bottomrule
\end{tabular*}
\vspace{-10pt}
\end{table*}
 
\vspace{4pt}  \noindent\textbf{Objective measures and suppression tests:}\label{sec:metrics}
We report improvement in SI-SNR (SI-SNRi)  relative to the input mixture \cite{leroux2019sdr}. We also report target preference (Picked): the percentage of outputs whose SI-SNR is higher against the target than against the sum of competing speech, which detects target confusion~\cite{zhao2022confusion}. We measure speaker similarity as the cosine similarity between Resemblyzer embeddings of output and target reference \cite{resemblyzer}; this evaluation encoder is independent of PVQE's enrollment encoder. Word accuracy (WAcc) is $100(1-\mathrm{WER})$ on Whisper \cite{radford2022whisper} transcripts. MCoRec uses corrected reference transcripts; the other corpora use Whisper transcripts of target references and thus measure transcription consistency. We report clip-median WAcc on synthetic tests and pooled word accuracy on meetings.
 
DNSMOS predicts speech (SIG), background (BAK), and overall (OVRL) quality from the output audio alone, without enrollment \cite{reddy2022dnsmosp835}; we use its calibration for personalized enhancement on all corpora and also report the DNSMOS P.808 overall estimate \cite{reddy2021dnsmos}.
 
We test preservation by supplying target-only speech and the target cues supported by each system. We count target-speaker over-suppression (TSOS) events lasting more than one second: contiguous flagged frames with excessive loss of the target's compressed spectral magnitude, using the detector and thresholds of \cite{eskimez2022personalized}. To measure changes in output level, we compute energy attenuation,
\vspace{-6pt}
\begin{equation}
A=10\log_{10}\frac{\sum_n x[n]^2}{\sum_n\hat s[n]^2}.
\vspace{-4pt}
\end{equation}
Unlike SI-SNR, this measure detects uniform gain loss. We report the fraction of target-only outputs with $|A|\leq3$ dB.
 
When the target is silent while other people talk nearby, an extractor should output silence. Following PVQE's target-absent evaluation \cite{parnamaa2024personalized}, we remove the target's audio and test two video conditions. A repeated still frame approximates a participant who is on camera but not speaking. A still mouth is itself a cue that the target is silent, so we also supply the target's video cyclically shifted within the excerpt: the mouth keeps moving but no longer matches the audio. Both audio-visual models receive the same audio and video, while AV-PVQE also receives enrollment.

\begin{table}[t]
\caption{AV-PVQE receives higher personalized P.835 ratings than AVASE on both meeting corpora. All six AV-PVQE--AVASE difference intervals exclude zero. Ratings use a 1--5 scale; the final column reports the difference and its 95\% bootstrap interval. \textbf{Bold} means are highest within each row.}
\label{tab:human}
\centering
\fontsize{9}{10.2}\selectfont
\setlength{\tabcolsep}{1.8pt}
\renewcommand{\arraystretch}{1.12}
\begin{tabular*}{\columnwidth}{@{\extracolsep{\fill}}lrrrr@{}}
\toprule
Scale & AVASE & PVQE & \shortstack{(Ours)\\AV-PVQE} & \shortstack{\(\Delta\) vs AVASE\\{[95\% CI]}}\\
\midrule
\multicolumn{5}{@{}l}{\textbf{AMI}}\\
SIG & 2.87 & \best{3.58} & 3.47 & \best{+.60 [.31, .90]}\\
BAK & 2.67 & 3.03 & \best{3.11} & \best{+.44 [.19, .69]}\\
OVRL & 2.11 & \best{2.74} & 2.68 & \best{+.57 [.32, .81]}\\
\midrule
\multicolumn{5}{@{}l}{\textbf{MCoRec}}\\
SIG & 1.65 & 2.39 & \best{2.63} & \best{+.98 [.74, 1.23]}\\
BAK & 1.28 & 1.84 & \best{1.94} & \best{+.66 [.49, .83]}\\
OVRL & 1.22 & 1.68 & \best{1.85} & \best{+.63 [.46, .80]}\\
\bottomrule
\end{tabular*}
\vspace{0pt}
\end{table}
 
\vspace{3pt}  \noindent\textbf{Personalized listening study: }
We run personalized ITU-T P.835 listening tests \cite{parnamaa2024personalized,naderi2021p835} for AMI and MCoRec on Prolific. Each trial plays 3--6s of the target's clean speech, a one-second marker, and the processed excerpt at its output level and without video, so listeners know whose voice to judge before rating SIG, BAK, and OVRL on five-point scales. Each listener rates 12 clips, four from each of the three systems, plus a trap and a gold clip as attention checks. Across both corpora, 395 screened listeners rated outputs from 107 excerpts per corpus, each processed by all three systems. A pilot run on LRS3 gave a 0.70 MOS standard deviation of per-clip OVRL differences; at 0.75 MOS, 107 paired excerpts detect 0.20 MOS at two-sided $\alpha=0.05$ with $\approx$80\% power. We report vote-weighted means and 95\% intervals from 50,000 bootstrap replicates that resample listeners and excerpts independently.
 
\vspace{-4pt}
\section{Results and discussion}
\vspace{-4pt}
\begin{figure}[t]
\centering
\begin{tikzpicture}
\begin{axis}[
  ybar, bar width=8pt,
  width=\columnwidth, height=3.5cm,
  symbolic x coords={Quieter,Equal,Louder}, xtick=data,
  enlarge x limits=0.25,
  xlabel={Target level relative to competing speaker},
  ylabel={Target preference (\%)},
  ymin=0, ymax=100, ytick={0,25,50,75,100},
  ymajorgrids, grid style={gray!20},
  tick label style={font=\footnotesize},
  label style={font=\footnotesize},
  legend style={at={(0.5,1.03)}, anchor=south, legend columns=3,
                draw=none, font=\footnotesize,
                /tikz/every even column/.append style={column sep=6pt}},
]
\addplot[fill=orange!70, draw=none]
  coordinates {(Quieter,93.92) (Equal,98.58) (Louder,98.88)};
\addplot[fill=black!40, draw=none]
  coordinates {(Quieter,6.25) (Equal,61.08) (Louder,98.63)};
\addplot[fill=blue!60!black, draw=none]
  coordinates {(Quieter,97.00) (Equal,99.42) (Louder,99.25)};
\legend{AVASE, PVQE, AV-PVQE (ours)}
\end{axis}
\end{tikzpicture}
\vspace{-12pt}
\caption{Target preference on LRS3 by target-to-interferer ratio; the equal group spans $\pm2$~dB. PVQE follows the louder voice; AV-PVQE selects the target at least 97\% of the time at every level.}
\label{fig:bylevel}
\vspace{-14pt}
\end{figure}
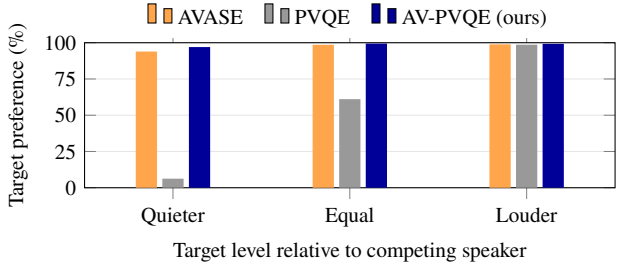

\subsection{Extraction on synthetic mixtures}\label{sec:synthetic}
\vspace{-2pt}
PVQE follows the louder voice (Fig.~\ref{fig:bylevel}). It selects the target in 98.63\% of mixtures where the target is louder and 6.25\% where it is quieter. PVQE selects the target in 61.08\% of equal-level mixtures. AV-PVQE selects the target at least 97\% of the time at every level, and its lead over AVASE is largest when the target is quieter (97.00\% vs.\ 93.92\%). On both synthetic benchmarks, AV-PVQE also improves SI-SNRi, word accuracy, and speaker similarity over AVASE (~\Cref{tab:extraction}).
 
AV-PVQE raises every DNSMOS estimate over AVASE on all four corpora. PVQE scores higher on nearly all, although its output is closer to the competing speaker in 46\% of LRS3 cases, so we compare quality with listeners who first hear the target's voice (\S\ref{sec:listening}).

\vspace{-4pt}
\subsection{Generalization to meetings}
AV-PVQE's margin over AVASE grows from synthetic mixtures to meetings: 0.31 dB SI-SNRi on LRS3, 1.84 dB on VoxCeleb2, 4.43 dB on AMI, and 6.58 dB on MTM (\Cref{tab:extraction}). Word accuracy rises from 50.14\% to 68.27\% on AMI, from 49.56\% to 67.65\% on MTM, and from 14.38\% to 25.38\% on MCoRec. AVASE learns reconstruction only from synthetic two-speaker mixtures, while AV-PVQE starts from a model trained to suppress noise, echo, and reverberation, conditions closer to those of meeting recordings.
 
\noindent AV-PVQE is fine-tuned on two-speaker mixtures, yet its lead over AVASE holds in every MTM speaker-count group, and its target-preference advantage grows from 9.2 points with two speakers to 18.6 and 18.2 points with three and four (\Cref{tab:speakers}). With four speakers, AVASE's target preference falls to 56.82\%, while AV-PVQE keeps 75.00\%.

\vspace{-4pt}
\subsection{Target preservation and rejection}\label{sec:preservation}
\vspace{-2pt}
Across 6,000 target-only LRS3 inputs, AV-PVQE produces seven long target-suppression events, against 142 for PVQE, and keeps 96.93\% of outputs within $\pm3$ dB of the input level (\Cref{tab:controls}).
 
With the target's audio removed and a still target image, AV-PVQE attenuates the competing speaker by 32.83 dB, against 2.25 dB for AVASE and 2.46 dB for PVQE, which has enrollment but no video. With shifted video, where the mouth keeps moving and AVASE has lip motion to use, AV-PVQE still attenuates competing speech by 18.28 dB, against 8.62 dB for AVASE.

\vspace{-4pt}
\subsection{Listener-rated quality}\label{sec:listening}
\vspace{-2pt}
Listeners rate AV-PVQE above AVASE on SIG, BAK, and OVRL for both AMI and MCoRec, and all six difference intervals exclude zero (\Cref{tab:human}). Overall quality improves by 0.57 MOS on AMI and 0.63 MOS on MCoRec.
 
The comparison with PVQE tests whether recovering the target costs perceptual quality. On AMI, AV-PVQE raises word accuracy from 18.51\% to 68.27\%, and its OVRL differs from PVQE's by $-0.06$ MOS (95\% CI: $[-0.24,0.12]$); on MCoRec the difference is $+0.17$ MOS ($[-0.01,0.35]$). Recovering the target therefore did not measurably reduce overall quality. 

\vspace{-5pt}
\section{Conclusion}
\vspace{-4pt}
AV-PVQE adapts a pretrained personalized speech enhancer for online audiovisual target-speaker extraction with no lookahead. It improves target recovery over PVQE while retaining similar mean overall-quality ratings, and outperforms AVASE in both target recovery and listener ratings. The separation gains over AVASE are larger on recorded meetings. Future work will measure device processing time and capture-to-output delay, and test whether joint enhancement and extraction training improves noise, echo, and reverberation suppression alongside extraction. We will also train separate models without video and from random initialization to assess the contributions of visual conditioning and enhancement pretraining.

\clearpage
\balance
\begingroup
\emergencystretch=1em
\bibliographystyle{IEEEbib}
\bibliography{references}
\endgroup
\section{Compliance with Ethical Standards}
We conducted the listening study as industry research through
Prolific. Participation was voluntary and compensated.
Participants provided informed consent, and ratings were
recorded under pseudonymous participant IDs.
\end{document}